\documentclass[prl, aps, twocolumn, superscriptaddress]{revtex4-2}

\usepackage[utf8x]{inputenc}

\usepackage{slashed}
\usepackage{graphicx}
\usepackage{subfigure}
\usepackage[usenames, dvipsnames]{color}
\usepackage{graphics}
\usepackage{hyperref}
\usepackage{bm}
\usepackage{fontenc}
\usepackage{amsmath}
\usepackage{color}
\usepackage{amsfonts}

\hypersetup{backref,
colorlinks=true,
linkcolor=blue,
linktoc=page,
citecolor=blue,
urlcolor=blue}

\everymath{\displaystyle}

\begin{document}

\title{Resonant axion-plasmon conversion in neutron star magnetospheres}

\author{H. Ter\c{c}as}
\email{hugo.tercas@isel.pt}
\affiliation{Instituto Superior de Engenharia de Lisboa, Instituto Politécnico de Lisboa, Lisboa, Portugal}
\affiliation{GoLP/Instituto de Plasmas e Fus\~ao Nuclear, Instituto Superior T\'ecnico, Lisboa, Portugal}

\author{J. T. Mendon\c{c}a}
\affiliation{GoLP/Instituto de Plasmas e Fus\~ao Nuclear, Instituto Superior T\'ecnico, Lisboa, Portugal}

\author{R. Bingham}
\affiliation{Central Laser Facility, STFC Rutherford Appleton Laboratory, Didcot, United Kingdom}
\affiliation{Department of Physics, University of Strathclyde, United Kingdom}

\begin{abstract}

Resonant axion-plasmon conversion in the magnetospheres of magnetars may substantially impact the landscape of dark-matter axion detection. This work explores how resonant axion-plasmon conversion, through a mechanism that is analogous to the Mikheyev-Smirnov-Wolfenstein (NSW) effect in neutrinos, modify the expected radio signals from axion-photon conversions observed on Earth. Critically, the resonant conversion radius lies within the region expected for axion-photon conversion, introducing a nonradiative power loss that diminishes the anticipated photon flux. Our analysis demonstrates that this effect can reduce radio telescope sensitivities, shifting them into regions excluded by previous experiments. These findings compel a reassessment of experimental constraints derived from radio signatures of axion-photon conversions and highlight the necessity of accounting for plasmon effects in astrophysical axion searches. The presented corrections provide critical insights for refining the detection strategies of future telescope-based dark matter axion experiments.

\end{abstract}
\maketitle

{\it Introduction.} The charge-parity (CP) violation remains one of the most fundamental, yet unresolved, issues in modern physics \cite{PhysRevLett.13.138, PhysRevLett.83.22, Christenson_1964, Cabibbo_1963, Gell-Mann1960, Pontecorvo_1958}. At the core of the so-called \textit{strong CP problem} is the anomalous electric dipole moment of the neutron \cite{PhysRevD.92.092003}. The Peccei-Quinn (PQ) mechanism provides a compelling resolution by promoting the CP-violating angle in the QCD Lagrangian to a dynamical field \cite{quinn_1977, kim_1978}, with its associated pseudo-Goldstone boson—the axion—emerging from the spontaneous breaking of a global U(1) symmetry \cite{weinberg_1978}.

Axions and axion-like particles (ALPs) are well-motivated candidates for dark matter, owing to their extremely small mass and weak coupling to standard model particles \cite{PhysRevLett.51.1415, PhysRevLett.104.041301}. Various experimental efforts, both terrestrial and astrophysical, have been dedicated to their detection \cite{PhysRevLett.94.121301, PhysRevLett.98.201801, PhysRevLett.118.261301}. However, the axion-photon coupling is weak, making their detection particularly challenging. Ground-based experiments such as PVLAS \cite{PhysRevLett.96.110406} have generated controversial results, while large-scale astrophysical searches, including CAST \cite{Collaboration2017}, ADMX \cite{asztalos_2001, asztalos_2010, admx_2018}, IAXO \cite{vogel_2015}, and MADMAX \cite{caldwell_2017}, have sought axion signals in different mass ranges. These searches rely on passive detection, motivating complementary laboratory-based strategies such as high-power laser interactions \cite{RevModPhys.84.1177}, plasma wakefield acceleration \cite{Mangles_2004, Geddes2004, Faure2004, adli_2018}, and direct axion production \cite{Burton_2018, Burton_2010, Burton_2016}.

Astrophysical settings provide natural environments for axion-photon conversion, particularly in strong magnetic fields \cite{Mirizzi_2008, Wang_2016, Masaki_2017, Hook_2018, Foster_2020, Witte_2021, Millar_2021,Buckley_2021,Dessert_2022, DeMiguel_2022, Seong_2024, Todarello_2025}. Axions propagating through the magnetospheres of neutron stars are known to resonantly convert into radio photons \cite{asseo_1980, bibber_1989, raffelt_1988, pshirkov_2009, srimoyee_2018}, a mechanism akin to the Mikheyev-Smirnov-Wolfenstein (MSW) effect in neutrino physics. Recent works have highlighted the role of plasma effects in modifying the conversion dynamics \cite{Hook_2018, Witte_2021}, but these studies focus solely on axion-photon interactions, neglecting the role of plasmons.
\begin{figure}[t!]
\includegraphics[width=\columnwidth]{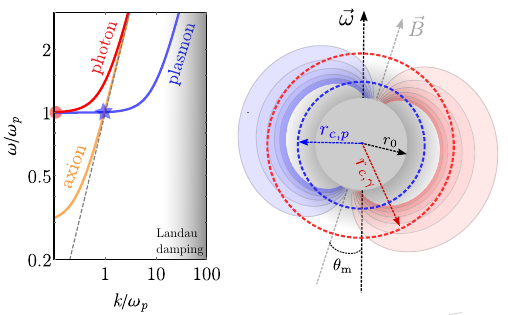}
\caption{(color online) {\bf Dispersion relation and mode conversion}. Left panel: Plasmon (solid blue line), photon (solid red) and axion (solid orange) lines, alongside with the light cone $\omega=k$ (gray/dashed line). The resonant, nonadiabatic axion-plasmon (adiabatic axion-photon) conversion point is marked by a star (circle). The Landau damped region, $k\gtrsim \omega_p/S_e$, is marked for reference. Right panel: schematics of a pulsar, with a misalignment angle $\theta_{\rm m}$ between the magnetic and rotation axis, displaying the resonant axion-plasmon conversion radius, $r_{c,p}$, and the adiabatic axion-photon conversion radius, $r_{c,\gamma}$. The effective conversion volume $4\pi/3 \left( r_{c,\gamma}^3-r_{c,p}^3\right)$ must be considered for the correct estimation of the photon flux density at Earth.}
\label{fig_dispersion}
\end{figure}
In this Letter, we demonstrate that axion-plasmon conversion plays a crucial role in neutron star magnetospheres, significantly altering the expected radio signal reaching Earth. The resonant axion-plasmon conversion occurs at a smaller radius than axion-photon conversion, effectively reducing the volume where detectable photons are generated. As a result, the total conversion efficiency is suppressed, leading to a dimming of the expected radio signal. This effect has been overlooked in previous studies, despite its critical impact on signal detectability.

Our analysis provides a quantitative framework to assess the suppression of the photon flux due to axion-plasmon conversion and its impact on telescope sensitivities. We show that, for typical neutron star conditions, this correction can significantly reduce the projected sensitivity of radio searches, potentially shifting viable parameter space into previously excluded regions. These results underscore the necessity of incorporating axion-plasmon interactions in astrophysical axion searches to obtain accurate constraints on axion properties. Future observational strategies must account for this effect to refine detection techniques and improve the interpretability of axion search experiments.

{\it Axion-plasmon dispersion relation.} The minimal electromagnetic theory accommodating the axion-photon coupling, while casting eventual QED effects in the plasma due to the strong magnetic fields of the magnetar, can be constructed as follows \citep{visinelli_2013, wilczek_1987, raffelt_1988, neves_2021} 
\begin{align}
\mathcal{L}&=-\frac{1}{4}F_{\mu\nu}F^{\mu\nu}-A_\mu J^\mu+\mathcal{L}_\varphi +\mathcal{L}_{\rm int}+\beta\left[\mathcal{F}^2+\frac{7}{4}\mathcal{G}^2\right]\nonumber\\&+\sigma\left[(\partial_\mu F^{\mu\nu})(\partial_\alpha F^{\alpha}_\nu)-F_{\mu\nu}\square  F^{\mu\nu}\right], 
\label{eq_lagrange}
\end{align}
where $F_{\mu \nu}=\partial_\mu A_\nu-\partial_\nu A_\mu$ is the electromagnetic (EM) tensor, $\square=\partial_t^2-\nabla^2$ is the d'Alembert operator, $\mathcal{F}=F_{\mu\nu}F^{\mu\nu}$, and $\mathcal{G}=F_{\mu\nu}\tilde F^{\mu\nu}$, with $\tilde F^{\mu \nu}=\epsilon^{\mu \nu \alpha \beta} F_{\alpha \beta}$ denoting the dual EM tensor, and $J^{\mu}$ is the four-current. The last two terms in Eq. \eqref{eq_lagrange} describe the Euler-Heisenberg (one-loop) QED corrections that are relevant for the strong magnetic fields of the magnetars. Here, $\beta=\alpha^2/(90m_e^4)$, with $\alpha=e^2/(4\pi)\simeq 1/137$ being the fine structure constant, and $\sigma=(2/15)\alpha/m_e^2$. Moreover, $\mathcal{L}_\varphi=\partial_\mu \varphi^*\partial^\mu \varphi-m_\varphi^2 \vert \varphi \vert^2$ is the axion Lagrangian (with $\varphi$ denoting the axion field). For the QCD axion, $m_\varphi=\sqrt{z}f_\pi m_\pi/f_\varphi$, where $z=m_u/m_d$ is the up/down mass ratio, and $f_{\varphi(\pi)}$ is the axion (pion) decay constant \cite{quinn_1977, weinberg_1978}. Upon integration of the anomalous of the axion-gluon triangle, one obtains $\mathcal{L}_{\rm int}=-(g/4)\varphi F_{\mu \nu}\tilde F^{\mu \nu}$, where $g$ is the axion-photon coupling. Although motivated for the QCD axion, the model in Eq. (\ref{eq_lagrange}) is valid for any axion-like particle (ALP). From Euler-Lagrange equations, one obtains the inhomogeneous Maxwell's equations \citep{tercas_2018, neves_2021, Lundin_2006} as
\begin{align}
\left(1+2\sigma\square\right)\partial_\mu F^{\mu\nu}&=J^\nu+2\beta\partial_\mu\left[ \mathcal{F} F^{\mu\nu}+\frac{7}{4}\mathcal{G}\tilde F^{\mu\nu}\right]\nonumber\\
&+g\partial_\mu\left(\varphi\tilde F^{\mu\nu}\right).
\end{align}
In the context of the axion-plasmon polariton, the only relevant equation is the  Poisson equation ($\nu=0$), which reads 
\begin{equation}
\bm\nabla \cdot \bm \left(\bm{\mathcal{E}}+g\varphi \bf B\right)=\rho, 
\label{eq_maxwell}
\end{equation}
where the effective electric reads
\begin{align}
\bm{\mathcal{E}}=\left[1+2\sigma\square +4\beta\left(E^2-B^2\right)+7\beta\left({\bf E} \cdot {\bf B}\right){\bf B}\right]{\bf E}. 
\label{eq_efield}
\end{align}
Moreover, variation with respect to $\varphi$ yields the Klein-Gordon equation describing the axion field
\begin{equation}
\left(\square +m_\varphi^2\right)\varphi= g {\bf E}\cdot {\bf B},
\label{eq_KG}
\end{equation}
The plasma density is $\rho=e(n_i-n_e)$, where $n_i$ and $n_e$ respectively represent the ion and electron densities. As we are interested in electron plasma waves only, we can assume the ions to be immobile. Thus, the equations governing the plasma electrons read
\begin{align}
\frac{\partial n_e}{\partial t}+\bm \nabla\cdot \left(n_e \mathbf{u}_e \right)&=0,
\label{eq_continuity}\\
\left(\frac{\partial}{\partial t}+\mathbf{u}_e\cdot \bm \nabla\right)\mathbf{u}_e&=-\frac{e}{\gamma_e m_e}\left(\mathbf{E}+{\bf u}\times {\bf B}\right)-\frac{\bm \nabla P_e}{n_e m_e},
\label{eq_force}
\end{align}
where $\gamma_e=(1-u_e^2)^{-1/2}$ is the Lorentz factor. In the following, and without any loss of generality, we will consider the plasma electrons to be nonrelativistic ($\gamma_e\simeq 1$). At first, we are interested in describing the electrostatic perturbations along a static, homogeneous magnetic field ${\bf B}=B_0 {\bf e}_z$. We retain first-order terms in the perturbations only, taking $n_e= n_0 + \tilde n_e$ and $\varphi=\tilde \varphi$ (neglecting the presence of a vev, $\varphi_0=0$). Further, we assume the compressions in the electron fluid to be adiabatic, $\bm \nabla \tilde P_e\simeq \gamma T_e\bm  \nabla \tilde n_e$, with $\gamma$ being the adiabatic index. Putting Eqs. \eqref{eq_maxwell}$-$\eqref{eq_force} together, we obtain the following eigenvalue problem
\begin{align}
\left(\omega^2-\frac{\omega_p^2}{\mathcal{D}(\omega,k)}+S_{e}^2k^2\right)\tilde n_e&-\frac{gB_0\omega_p}{\mathcal{D}(\omega,k)}\tilde \psi=0,\\
\left(\omega^2-k^2+m_\varphi^2+\frac{g^2 B_0^2}{\mathcal{D}(\omega,k)} \right) \tilde\psi&-gB_0\omega_p\tilde n_e=0,
\label{eq_dispQED}
\end{align}
where $\tilde \psi=ikm_e\tilde\varphi/e$ is an auxiliary field, $\omega_p=\sqrt{e^2 n_0/m_e}$ is the plasma frequency and $S_e=\sqrt{\gamma T_e/m_e}$ is the electron thermal speed. The quantity 
\begin{equation}
\mathcal{D}(\omega,k)=1+3\beta B_0^2-2\sigma(\omega^2-k^2)
\end{equation} 
accounts for the QED effects, which relevance strongly depends on the energy scale at which the axion-plasmon mixing takes place. Considering resonant processes only ($\omega \simeq \omega_p$), relativistic axions may be produced by plasmons if $m_\varphi\ll \omega_p$, making resonance possible near the light cone $\omega_p\simeq k$, for which $\mathcal{D}(\omega, k)\simeq 1+3\beta B_0^2$. On the other hand, plasmons resonantly convert into non-relativistic axions provided the condition $m_\varphi\simeq \omega_p$, and thus $\mathcal{D}(\omega, k)\simeq 1+3\beta B_0^2-2\sigma \omega_p^2.$ For the typical conditions found at the polar caps of magnetars, $B_0\sim 10^9$ T and $\omega_p\sim 10^{-3}$ eV, which leads $\mathcal{D}(\omega, k)\simeq 1$ for both cases, allowing us to safely neglect QED effects in the resonant plasmon-axion conversion. Although instructive, this discussion could be avoided by noticing that the dispersive QED term $\sim \sigma$ is only relevant at the energy scales $k\sim m_e$, which lie well above the plasma frequency ($\omega_p\ll m_e$). In the absence of QED effects, $\mathcal{D}(\omega, k)\sim 1$, Eqs. \eqref{eq_dispQED} can be recast in a more compact form as $\bm\nabla\cdot(\epsilon(k, \omega){\bf E})=0 $, where
\begin{equation}
\epsilon(k, \omega)=  1-\frac{\omega_e^2}{\omega^2}-\frac{\Omega^4}{\omega^2(\omega^2-\omega_\varphi^2)}
\label{eq_dispersion}
\end{equation}
is the dielectric permittivity, $\omega_e^2=\omega_p^2+S_e^2k^2$ is the plasmon dispersion, $\omega_\varphi^2=M_\varphi^2+k^2$ is the axion dispersion, and $M_\varphi=\sqrt{m_\varphi^2+g^2B_0^2}$ is the axion effective mass in the plasma. Here, $\Omega=\sqrt{gB_0 \omega_p}$ is the axion-plasmon coupling parameter (Rabi frequency), which reads 
\begin{equation}
\Omega\simeq 2\pi\times (1.2 {~\rm MHz})\sqrt{\frac{g}{ {~\rm 10^{-12} GeV}^{-1}}\frac{B_0}{ {\rm 10^{10}T}}\frac{\omega_p}{\rm  GHz}}
\end{equation}
Equation \eqref{eq_dispersion} yields the secular equation $\left(\omega^2-\omega_\varphi^2\right)(\omega^2-\omega_e^2)=\Omega^4$, containing the lower (L) and (U) polariton modes \cite{tercas_2018}
\begin{equation}
\omega_{\rm L(U)}^2=\frac{1}{2}\left[ \omega_\varphi^2 +\omega_e^2 \pm \sqrt{(\omega_e^2-\omega_\varphi^2)^2+4\Omega^4}\right].
\label{eq_polariton}
\end{equation}  
Given the smallness of the coupling, $\Omega\ll \omega_p$, we can linearize the secular equation around the resonant mode $k_c$ satisfying the condition $\omega_\varphi=\omega_e\simeq \omega_p$ (see Fig. \ref{fig_dispersion}) as
\begin{equation}
(\omega-\omega_\varphi)(\omega-\omega_e)\simeq \frac{\Omega^4}{4\omega_p^2}=\frac{g^2B_0^2}{4}.
\label{eq_RWA}
\end{equation}
Since the coupling $g B_0$ is small comparing to the bare frequencies $\omega_e$ and $\omega_\varphi$, the two modes can propagate independently over most of the magnetosphere, except at the resonance point $(r_c, k_c)$. In the magnetosphere of a neutron star (NS), both the magnetic field and the electronic density are inhomogeneous, following the Goldreich-Julian (GJ) profiles \cite{GJ_1969, Hook_2018, Witte_2021}. The radial component of the magnetic field reads
\begin{align}
B(r)&=B_0\left(\frac{r_0}{r}\right)^3 f, \nonumber \quad {\rm with}\\
f \equiv f(\theta,\theta_{\rm m},\psi)  = &\cos\theta_{\rm m}\cos\theta +\sin\theta_{\rm m}\sin\theta\cos\psi
\label{eq_GJ}.
\end{align}
Here, $r_0$ is the NS radius of the, $B_0$ is the magnetic field at the pole caps, $\theta_m$ is the misalignment angle between ${\bf B}$ and the NS angular velocity ${\bm\omega}_{\rm NS}$, $\psi= \phi-\omega_{\rm NS} t$ and $\theta$ and $\phi$ are the usual polar and azimuthal angles. The radial electron density is then given by $n(r)=n_0 (r_0/r)^3$, with $n_0= 2\pi B_0/(e\tau)$ being the electronic density at the NS surface and $\tau=2\pi/\omega_{\rm NS}$ being the rotation period. Although including angular corrections is straightforward, we ignore them as we are interested in order of magnitudes estimates. Since these variations are smooth at the scale of $1/k_c$, we can  assume that, outside the resonance points, the two independent modes are given by $\omega=\omega_\varphi(r, k)$ and $\omega=\omega_e(r,k)$. At the resonance point $r=r_c$, we have $\omega_\varphi(r_c,k_c)=\omega_e(r_c, k_c)$. To linearize around this point, we can use the prescription  
\begin{equation}
r=r_c+\xi, \quad k=k_c+\delta.
\end{equation}
\label{eq_variation}
Upon substitution in Eq. \eqref{eq_RWA}, and using the conversion condition, we obtain \cite{Cairns_1982, Cairns_1983b}
\begin{equation}
\left(a_\varphi \delta+b_\varphi\xi \right)\left(a_e \delta+b_e \xi \right)\simeq \frac{g^2 B_c^2}{4},
\label{eq_RWA2}
\end{equation}
where $B_c=B(r=r_c)$, and $a_\alpha=\partial_{k}\omega_\alpha\vert_{k=k_c}$ and $b_\alpha=\partial_r\omega_\alpha\vert_{r=r_c}$. Provided the identification $\delta=-i \partial_\xi$, and noticing that the wave amplitude varies as $\Phi(r)\sim \exp(ik_c r)$, Eq. \eqref{eq_RWA2} can be recast in the following form
\begin{equation}
\begin{array}{c}
-a_\varphi a_e\frac{\partial^2\Phi}{\partial \xi^2}-i\left(a_\varphi b_e-a_e b_\varphi\right) \xi \frac{\partial\Phi}{\partial \xi}\\ +\left(a_\varphi b_e\xi^2-\frac{g^2 B_c^2}{4}\right)\Phi=0.
\end{array}
\label{eq_RWA3}
\end{equation}
Away from the resonance point $r=r_c$, the solution can be taken as $\Phi\sim \exp\left[i\int \delta_\pm(\xi) d\xi\right]$, where 
\begin{widetext}
\begin{equation}
\delta_\pm(\xi)=\frac{1}{2a_\varphi a_e}\left[-\left(a_\varphi b_e+b_\varphi a_e\right)\xi \pm \sqrt{\left(a_\varphi b_e-b_\varphi a_e\right)^2\xi^2+a_\varphi a_e g^2 B_c^2} \right].
\end{equation}
\end{widetext}
In order to compute the plasmon-axion conversion probability, we must solve Eq. \eqref{eq_RWA3} exactly. For that, we proceed to a unitary transformation
\begin{equation}
\Psi(\xi)=\exp\left(i\frac{K\xi^2}{2}\right)\Phi(\xi), \quad K=-\frac{a_\varphi b_e+a_e b_\varphi}{2a_\varphi a_e},
\label{eq_transformation}
\end{equation}
and rescale the spatial coordinate, $\zeta= \sqrt{\left(a_\varphi b_e-a_e b_\varphi \right)/(a_\varphi a_e)} \exp(i 3\pi/4)\xi$, to obtain
\begin{equation}
\frac{\partial ^2\Psi}{\partial \zeta^2}+\left[\frac{ig^2B_c^2}{4\left(a_\varphi b_e- a_eb_\varphi\right)} + \frac{a_\varphi b_e+a_eb_\varphi}{2\left(a_\varphi b_e- a_eb_\varphi\right)}  -\frac{\zeta^2}{4} \right]\Psi=0.
\label{eq_RWA4}
\end{equation}
Finally, we notice that for a non-relativistic plasma, $S_e\ll 1$, so we can neglect the $k$ dependence on the plasma dispersion and safely take $a_e=0$. In that case, Eq. \eqref{eq_RWA4} yields the Weber's equation
\begin{equation}
\frac{\partial ^2\Psi}{\partial \zeta^2}+\left(\nu+\frac{1}{2}-\frac{\zeta^2}{4}\right)\Psi=0,
\label{eq_Weber}
\end{equation}
with $\nu=i g^2 B_c^2/4(a_\varphi b_e)$. The solution to Eq. \eqref{eq_Weber} is the well-known parabolic cylinder function $D_\nu(\zeta)$ \cite{Whittaker_1996}. For definiteness, we consider that $\Phi(\xi)$ describes an incoming axion and an outcoming plasmon in what follows. We notice, however, that this prescription is arbitrary, and reverting it amounts to change the definitions of the coefficients $a_\alpha$ and $b_\alpha$. With this prescription, we observe that, for $\xi<0$, we may write
\begin{equation}
\zeta=\left(\frac{a_\varphi b_e-a_e b_\varphi}{a_\varphi a_e}\right)^{1/2}\exp\left(i\frac{3\pi}{4}\right)\vert \xi\vert \exp\left(-i\pi\right)
\end{equation}
and expand the solution as $D_\nu^<(\zeta)\sim \zeta^\nu \exp(-\zeta^2/4)$, what yields, together with the transformation in \eqref{eq_transformation}, the following form for the incoming axion 
\begin{equation}
\Phi_<(\xi) \sim \exp\left(-\frac{i b_\varphi}{2 a_\varphi}\xi^2\right).
\label{eq_incoming}
\end{equation}
Conversely, in the outwards region, $\xi >0$, we have 
\begin{equation}
\zeta=\left(\frac{a_\varphi b_e-a_e b_\varphi}{a_\varphi a_e}\right)^{1/2}\xi \exp\left(i\frac{3\pi}{4}\right),
\end{equation}
so ${\rm arg} (\zeta)=3\pi/4$, yielding
\begin{equation}
\begin{array}{c}
D_\nu^>(\zeta)\sim \zeta^\nu \exp(-\zeta^2/4)	
-\frac{\sqrt{2\pi}}{(-\nu-1)!}\frac{\exp\left(i\nu \pi+\zeta^2/4\right)}{\zeta^{(\nu+1)}},
\end{array}
\label{eq_outcoming}
\end{equation}
and an similar expression for $\Phi_>(\xi)$. The first term of Eq. \eqref{eq_outcoming} represents the axion mode, while the second represents the plasmon mode. The conversion amplitude at the resonance $r_c$ is thus given by the ration of the axion amplitudes at $r<r_c$ ($\xi <0$) and $r >r_c$ ($\xi>0$), $\vert \exp(i\nu \pi) \vert$. The square of this quantity gives the axion transmission coefficient $T=\exp(-2\pi g^2 B_c^2/(4a_\varphi b_e))$. From energy conversion arguments, the plasmon-axion conversion amplitude, our quantity of interest, is therefore given by
\begin{equation}
P=\vert 1-T\vert =\left\vert 1-\exp\left(\frac{\pi g^2 B_c^2}{2 a_\varphi b_e}\right)\right\vert.
\end{equation}
For the conditions of a neutron star magnetosphere, the GJ profiles in Eq. \eqref{eq_GJ} yield
\begin{equation}
\begin{array}{c}
a_\varphi=\frac{k_c}{\omega_\varphi(r_c,k_c)}\simeq \left(1-\frac{m_\varphi^2+g^2B_0^2\frac{r_0^6}{r_c^6} f^2}{\omega_p^2}\right)^{1/2}\\
b_e=\frac{\partial \omega_e}{\partial r} \simeq -\frac{3}{2}\omega_p\frac{r_0^{3/2}f^{1/2}}{r_c^{5/2}},
\end{array}
\end{equation}
where we have used $k_c\simeq \sqrt{\omega_p^2-m_\varphi^2-g^2 B_c^2}$. The maximum probability for the plasmon-axion conversion occurs for at the radius $r_{c,p}$ such that $P=1$ and, therefore,
\begin{equation}
r_{c,p} \simeq r_0\left(\frac{4\omega_p^2-g^2 B_0^2}{4m_\varphi^2}\right)^{1/6}f^{1/3}.
\label{eq_rcp}
\end{equation}
\begin{figure}[t!]
\includegraphics[width=\columnwidth]{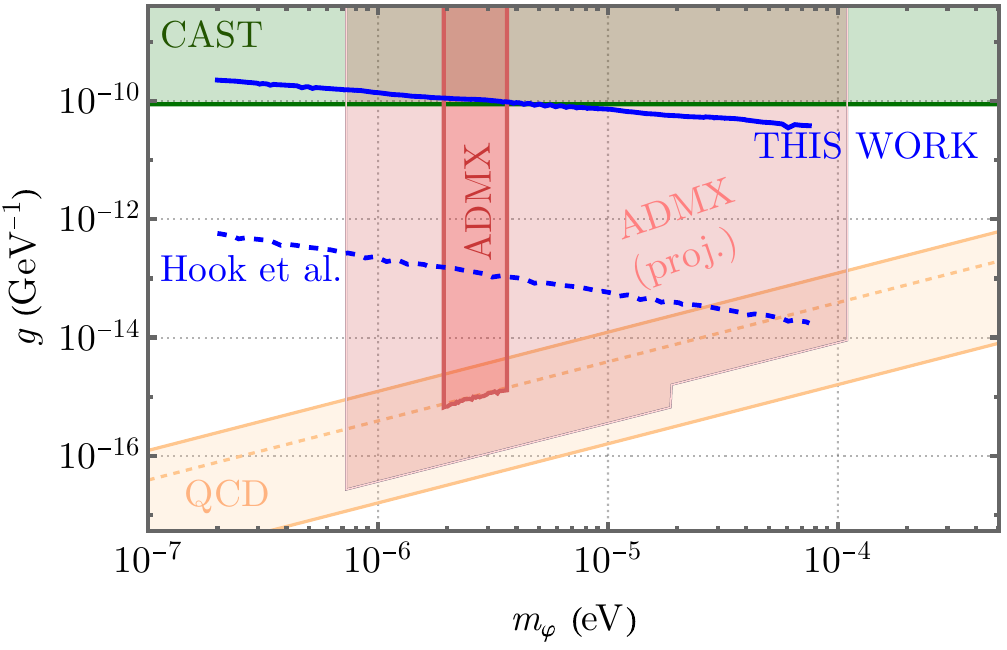}
\caption{(color online). {\bf Impact of axion-plasmon resonance on the sensitivity of photon signals emitted from neutron stars.} The dashed line corresponds to project sensitivity for radio signals observations due to axion-photon conversion in the neutron star SGR J1745-2900 (spike configuration, $\theta=90^\circ$, $\theta_m=10^\circ$, $\tau =100$ h), as calculated by Hook at al. \cite{Hook_2018}. Our estimates for the dimmed sensitivity, as consequence of the radiated power reduction due to resonant axion-plasmon conversion in Eq. \eqref{eq_corrected}, is depicted by the solid line. }
\label{fig_sensitivity}
\end{figure}
This should be compared with the conversion radius $r_{c,\gamma}$ where an axion decays into a O$-$mode (${\bf E}\parallel {\bf B}$) photon \cite{Hook_2018}, $r_{c,\gamma} \simeq r_0 \left(\omega_p/m_\varphi \right)^{1/3} f^{1/3}$. As we show below, this may drastically impact the estimated radiation power due to dark-matter axions converting into photons. The radiated power per solid angle for adiabatic (in opposition to resonant) photon conversion is given by \cite{Hook_2018}
\begin{equation}
\frac{d\mathcal{P}_\gamma}{d\Omega} \simeq \frac{1}{2} \frac{g^2 B_{c,\gamma}^2}{m_\varphi}\rho_{c,\gamma}^{\rm (DM)}V_{c,\gamma},
\end{equation}  
with $V_c=(4/3)\pi r_{c,\gamma}^3$ is the resonant conversion volume and $\rho_{c,\gamma}^{\rm (DM)}$ is the dark-matter density calculated at the conversion radius $r_{c,\gamma}$. To this quantity, one must subtract the volume leading to resonant axion-plasmon conversion, taking place at the radius $r_{c,p}<r_{c,\gamma}$, which consists in a nonradiative power loss. Provided the relations imparted by Eq. \eqref{eq_rcp}, the corrected radiated power reads,
\begin{equation}
\frac{d\mathcal{P}_\gamma^{\rm(corr)}}{d\Omega}\simeq \frac{d\mathcal{P}_\gamma}{d\Omega} \left[1-\left(1-\frac{g^2B_0^2}{4\omega_p^2}\right)^{1/2}\right].
\label{eq_corrected}
\end{equation}
The latter, in turn, corrects the telescopes flux density at Earth, $S_\gamma=(d\mathcal{P}_\gamma/d\Omega)/d^2\Delta\nu$, with $\Delta\nu$ denoting the bandwidth. To project the sensitivity of the radio signals, we have to compare $S_\gamma$ agains the minimal flux to which telescopes are responsive, $S_{\rm \min}={\rm SNR}_{\rm min}{\rm SEFD}/\sqrt{n_{\rm pol}\Delta\nu \tau_{\rm obs}}$, where ${\rm SNR}_{\rm min}$ is the minimal signal-to-noise ratio, SEFD is the system-equivalent flux density, $\tau_{\rm obs}$ is the observation time, and $n_{\rm pol}$ is the number of photon polarizations (for numerical evaluations, we choose $n_{\rm pol}=2$. In Fig. \eqref{fig_sensitivity}, we illustrate the corrections to the sensitivity to DM axion-photon conversion in the neutron star SGR J1745-2900, as calculated in Ref. \cite{Hook_2018}. Considering resonant axion-plasmon around the typical value $\omega_p=10^{-3}$ eV, we observe that the sensitivity can be drastically reduced, eventually falling within previously excluded regions. We believe that this effect may significantly alter the landscape of experiments based on radio telescope searches in the years to come.

\emph{Conclusion}. The resonant conversion of axions into plasmons in magnetar magnetospheres introduces a significant correction to the expected photon signals detectable on Earth. Our analysis shows that the radius at which axion-plasmon conversion occurs is smaller than the axion-photon conversion radius, effectively reducing the volume where detectable photons can emerge. This discrepancy, proportional to the difference in the respective conversion volumes, necessitates a reevaluation of experimental expectations based solely on axion-photon conversion mechanisms. The results presented here underscore the importance of accounting for axion-plasmon interactions in strongly magnetized environments, as they could hinder the photon flux reaching Earth and subsequently affect the interpretability of experimental data searching for axion signals. Future observational strategies must incorporate these findings to accurately constrain axion properties and refine detection techniques.    
\par

\emph{Acknowledgments}. The authors acknowledge FCT - Funda\c{c}\~{a}o da Ci\^{e}ncia e Tecnologia (Portugal) through the Grant No CEECIND/00401/2018 and Project No PTDC/FIS-OUT/3882/2020. This paper is based upon work from COST Action COSMIC WISPers CA21106, supported by COST (European Cooperation in Science and Technology). H.T. warmly thanks Thomas Grismayer for a critical reading of the manuscript.

\bibliographystyle{apsrev4-2}
\bibliography{references.bib}

\end{document}